\documentclass[10pt,twocolumn,letterpaper]{article}

\usepackage{wacv}
\usepackage{times}
\usepackage{epsfig}
\usepackage{graphicx}
\usepackage{amsmath}
\usepackage{amssymb}
\usepackage{booktabs}
\usepackage{booktabs}
\usepackage{color}
\usepackage{tabularx}

\newcommand\Tstrut{\rule{0pt}{2.0ex}}         % = `top' strut
\newcommand\Bstrut{\rule[-0.9ex]{0pt}{0pt}}   % = `bottom' strut

\def\wacvPaperID{309} % Enter the WACV Paper ID here

\wacvapplicationstrack % Uncomment this line if you are submitting to the Applications Track.

\wacvfinalcopy % *** Uncomment this line for the final submission

\ifwacvfinal
\usepackage[breaklinks=true,bookmarks=false]{hyperref}
\else
\usepackage[pagebackref=true,breaklinks=true,colorlinks,bookmarks=false]{hyperref}
\fi

\begin{document}

%%%%%%%%% TITLE
\title{DSFormer: A Dual-domain Self-supervised Transformer for Accelerated Multi-contrast MRI Reconstruction}

\author{Bo Zhou$^{1,3}$ { } { } Neel Dey$^{2,3}$ { } { } Jo Schlemper$^{3}$ { } { } Seyed Sadegh Mohseni Salehi$^{3}$ \\
Chi Liu$^{1}$ { } { } James S. Duncan$^{1}$ { } { } Michal Sofka$^{3}$\\
$^1$Yale University, New Haven, CT, USA \\
$^2$New York University, New York, NY, USA \\
$^3$Hyperfine Research, Guilford, CT, USA 
% {\tt\small bo.zhou@yale.edu}
% \and
% Neel Dey\\
% New York University\\
% New York, NY, USA\\
}

\maketitle
\thispagestyle{empty}

%%%%%%%%% ABSTRACT
\begin{abstract}
Multi-contrast MRI (MC-MRI) captures multiple complementary imaging modalities to aid in radiological decision-making. Given the need for lowering the time cost of multiple acquisitions, current deep accelerated MRI reconstruction networks focus on exploiting the redundancy between multiple contrasts. However, existing works are largely supervised with paired data and/or prohibitively expensive fully-sampled MRI sequences. Further, reconstruction networks typically rely on convolutional architectures which are limited in their capacity to model long-range interactions and may lead to suboptimal recovery of fine anatomical detail. To these ends, we present a dual-domain self-supervised transformer (DSFormer) for accelerated MC-MRI reconstruction. DSFormer develops a deep conditional cascade transformer (DCCT) consisting of cascaded Swin transformer reconstruction networks (SwinRN) trained under two deep conditioning strategies to enable MC-MRI information sharing. We further use a dual-domain (image and k-space) self-supervised learning strategy for DCCT to alleviate the costs of acquiring fully sampled training data. DSFormer generates high-fidelity reconstructions which outperform current fully-supervised baselines. Moreover, we find that DSFormer achieves nearly the same performance when trained either with full supervision or with the proposed self-supervision. 
\end{abstract}

%------------------------------------------------------------------------
\section{Introduction}
Diagnosticians often capture a series of multi-contrast magnetic resonance images (MC-MRI) of a single subject to acquire complementary tissue information towards more accurate and comprehensive radiological evaluation~\cite{menze2014multimodal,bakas2017advancing}. However, due to physical constraints, MRI intrinsically requires prolonged acquisition which often leads to patient discomfort and the accumulation of motion artifacts and system imperfections in the image that obfuscate biomedically-relevant anatomical detail. These limitations have lead to immense interest in accelerated methods that can reconstruct high-fidelity and artifact-free images from fewer (undersampled) frequency-domain (\textit{k-space}) MRI measurements and reduced scan time.

While the inverse Fourier transform can reconstruct images from fewer k-space measurements, it comes at the cost of strong aliasing and blurring effects in the reconstruction and has thus motivated works which exploit transform-domain data priors to achieve higher quality reconstructions with fewer artifacts~\cite{lustig2007sparse,donoho2009message,otazo2010combination}. However, these methods may still yield blurred and sub-clinical reconstructions and are generally slow and hyperparameter-sensitive as they are based on iterative \textit{instance-specific} optimization. More recently, deep MRI reconstruction networks have greatly improved MRI reconstruction fidelity under high undersampling rates with prediction times on the order of seconds~\cite{wang2016accelerating,schlemper2017deep,qin2018convolutional,hammernik2018learning,aggarwal2018modl,eo2018kiki,zhu2018image,zbontar2018fastmri,bien2018deep,zhang2019reducing,wang2020neural,yaman2020self,wei2020tuning,pineda2020active,liu2021universal,guo2021over,singh2020joint,zhou2020dudornet,ran2020md}. 

\begin{figure*}[tb!]
\centering
\includegraphics[width=0.98\textwidth]{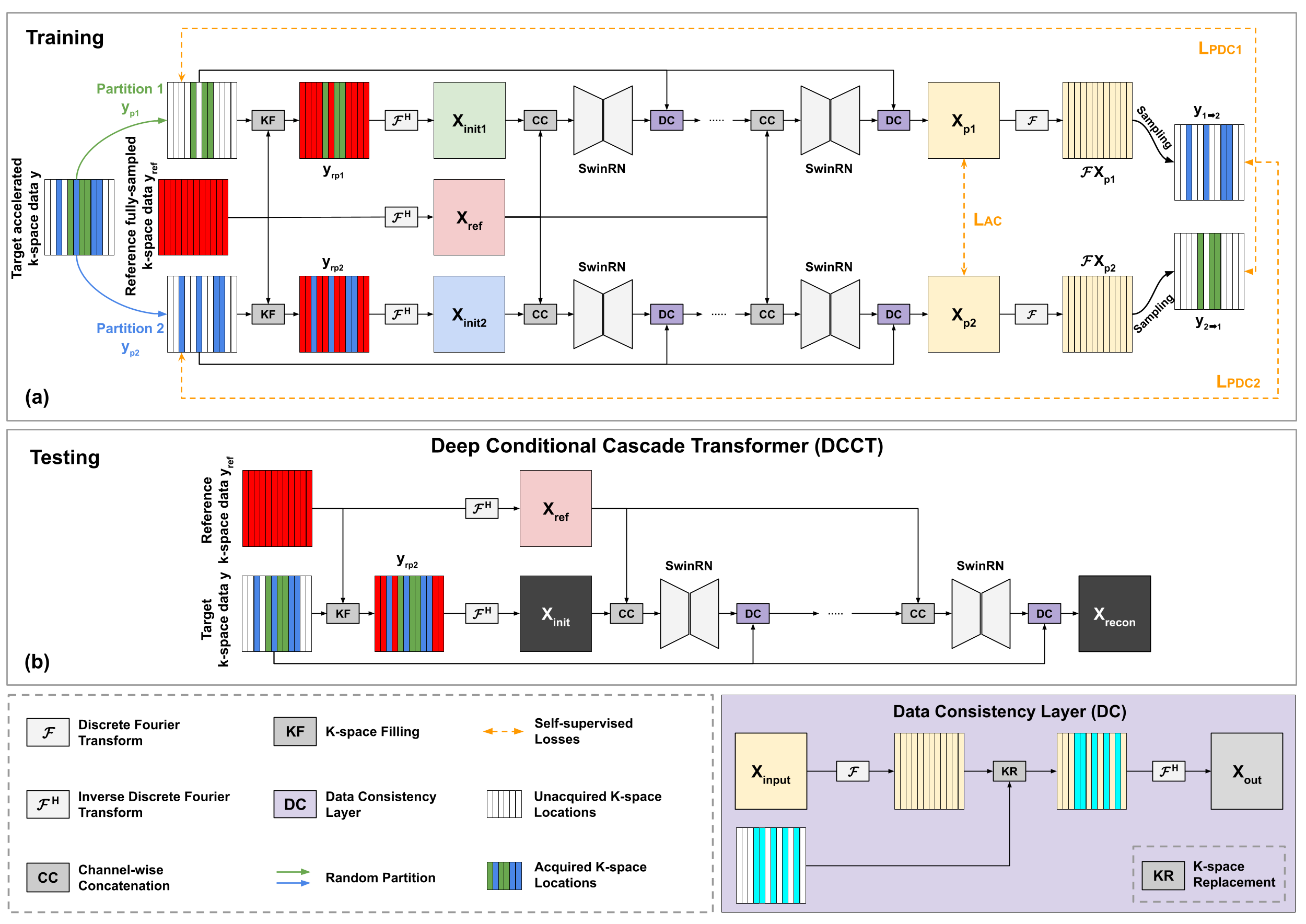}
\caption{An overview of the \textbf{D}ual-domain \textbf{S}elf-supervised Trans\textbf{Former} (\textbf{DSFormer}). The Deep Conditional Cascade Transformer (DCCT) is trained in a self-supervised fashion with randomly partitioned undersampled k-space data sets fed into DCCT in parallel. The partition data consistency loss ($\mathcal{L}_{PDC}$) and appearance consistency loss ($\mathcal{L}_{AC}$) are used for self-supervised training. At testing, arbitrarily undersampled data is reconstructed by DCCT. SwinRN (Fig. \ref{fig:network}) is used as the backbone network for DCCT.}
\label{fig:pipline}
\end{figure*}

However, these works typically achieve their strong results via supervised training on ground-truth \textit{fully-sampled} images and/or k-space target data, which is often practically infeasible in both time and cost to acquire. Recently, \textit{self-supervised} reconstruction frameworks have emerged requiring only undersampled k-space data~\cite{wang2020neural,yaman2020self,hu2021self}, yet their performance remains upper-bounded by full supervision. Further, whether supervised or self-supervised, the aforementioned works largely focus on single contrast MRI acceleration, whereas most diagnostics require MC-MRI to visualize disparate anatomical characteristics. Fortunately, in MC-MRI reconstruction, fully sampled MRI modalities requiring shorter acquisitions can be used as a reference to guide target modalities that require longer acquisitions via methods which inject a fully-sampled reference modality as an extra input channel into a reconstruction network~\cite{xiang2018ultra,sun2019deep,dar2020prior,zhou2020dudornet,liu2021deep,liu2021regularization}. 

While these MC-MRI methods achieve excellent reconstructions, they have the following major limitations. First, previous MC-MRI reconstruction methods operate directly on the undersampled target MRI image as input (reconstructed via zero-padding and the inverse Fourier transform) and thus suffer from severe aliasing in their starting point. Second, current MC-MRI reconstruction networks ubiquitously employ convolutional architectures~\cite{liu2021deep,xiang2018ultra,dar2020prior,ronneberger2015u}, such as U-shaped network designs~\cite{ronneberger2015u} and sequential convolutional layers with residual connections~\cite{liu2021deep,schlemper2017deep}, both of which are limited in modeling long-range interactions and may recover reduced fine image detail due to the lack of \textit{non-local} contextual information. Third, existing MC-MRI reconstruction methods require fully-supervised and fully-sampled training data from large-scale paired data, which is prohibitively expensive to obtain. The fully-sampled data of target contrasts demanding longer acquisitions are also prone to motion and other accumulating errors. Therefore, self-supervised learning operating on undersampled data with shorter acquisitions would be less susceptible to non-ideal imaging conditions.

\begin{figure}[!htb]
\centering
\includegraphics[width=0.48\textwidth]{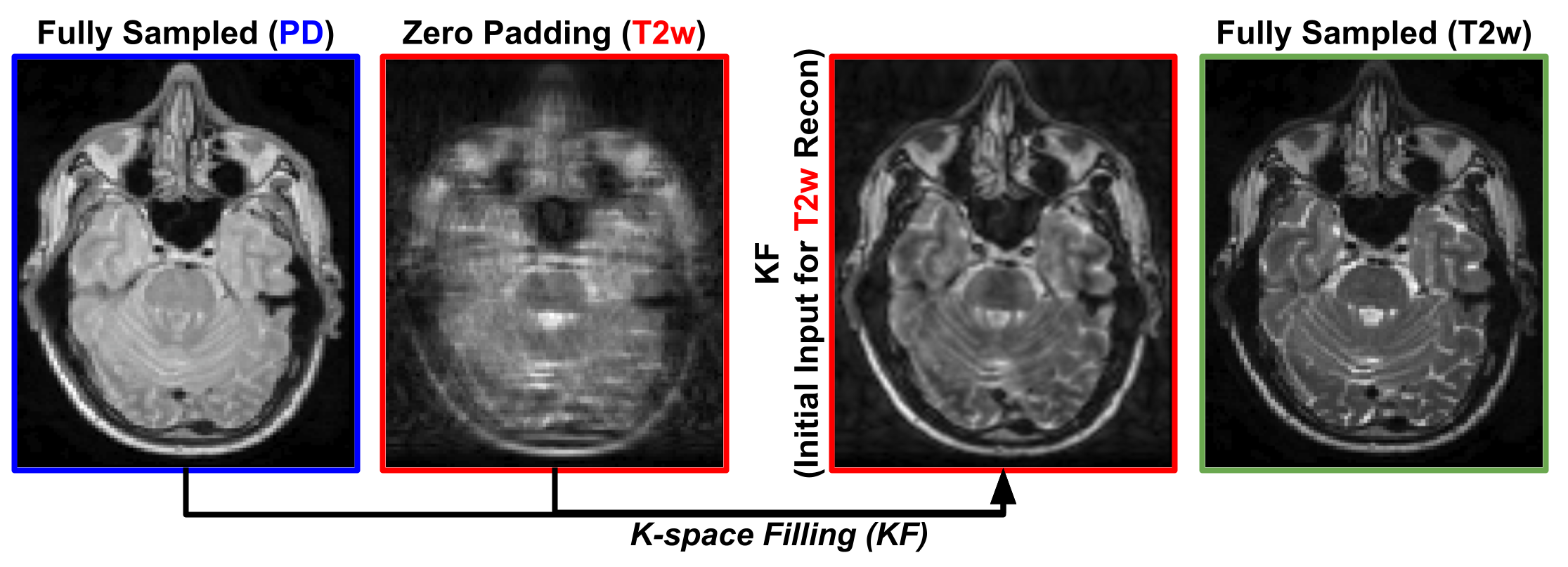}
\caption{K-space filling (KF) conditioning for $\times 4$ accelerated T2w reconstruction. Undersampled T2w data is combined with fully sampled PD data via KF to generate an initial DCCT input generating much fewer artifacts as compared to zero-padding.}
\label{fig:kfc}
\end{figure}

To these ends, we present DSFormer, a dual-domain self-supervised transformer for accelerated MC-MRI reconstruction, with the following contributions:
\begin{enumerate}
\item \textit{Multi-contrast information sharing.} We develop a deep MC-MRI conditioning method for efficient usage of multi-contrast information in MC-MRI reconstruction. Briefly, as opposed to the zero-padded and aliased initial reconstruction used in most works, our framework leverages fully-sampled \textit{reference} MRI data by grafting its k-space data into the unacquired k-space bins of the undersampled/accelerated \textit{target} modality, whose inversion provides a sharp, de-aliased, and anatomically-correct starting point for the network to operate on (Figure \ref{fig:kfc}). To further reinforce reference information in the undersampled reconstruction, we also channel-wise concatenate the reference MRI alongside the network inputs.  
%First, we leverage the fully sampled reference MRI data by grafting in its k-space data into the un-acquired k-space of the accelerated target MRI, providing an initial de-aliasing image for reconstruction network input (Figure \ref{fig:kfc}). Then, to reinforce sharing the reference MRI information, we also channel-wise concatenate the reference MRI with the network inputs. 
\item \textit{Vision Transformers for MRI reconstruction}. Inspired by recent advances in vision transformers showing improved image restoration over CNNs~\cite{wang2021uformer,chen2021pre,liang2021swinir,liu2021swin} by using non-local processing to recover fine detail, we develop a Swin transformer Reconstruction Network (SwinRN) to be used as a backbone in a cascaded framework. By combining MC-MRI conditioning with SwinRN-cascades, we propose a Deep Conditional Cascade Transformer (DCCT) for high-fidelity MRI reconstruction. 
%Previous reconstruction networks rely on either CNN consisting of sequential convolutional layers \cite{liu2021deep} or based on U-shape design \cite{xiang2018ultra,dar2020prior,ronneberger2015u}, which are sub-optimal for fine detail restoration. Inspired by the recent advances in image transformer for image restorations \cite{wang2021uformer,chen2021pre,liang2021swinir,liu2021swin}, we propose a Swin Transformer Reconstruction Network (SwinRN), and is used as the backbone network in our cascade framework. Combining the aforementioned MC-MRI deep conditioning and a cascade of SwinIR, we build the Deep Conditional Cascade Transformer (DCCT) for our fast MC-MRI reconstruction task (Fig. \ref{fig:pipline}b).

\item \textit{Dual-domain self-supervised learning}. To train DCCT in a self-supervised fashion using only undersampled target MRI data, we further use a dual image and k-space domain self-supervised learning approach, achieving reconstruction quality comparable to fully supervised training.
\end{enumerate}

Extensive experiments on MC-MRI data with different acceleration protocols demonstrate that DSFormer trained with either full supervision or only self-supervision generates superior reconstructions over previous architectures and conditioning mechanisms with fully supervised training strategies. %Finally, DCCT trained with dual-domain self-supervision achieves comparable performance to training under full-supervision. 

%------------------------------------------------------------------------
\section{Related work}
\noindent\textbf{Fully-supervised MRI reconstruction.} Convolutional neural networks (CNNs) have been extensively studied to reconstruct images from undersampled k-space data. For example, Wang \etal \cite{wang2016accelerating} recover fully-sampled MRIs from undersampled acquisitions using supervised CNN training on paired data. Schlemper \etal \cite{schlemper2017deep,qin2018convolutional} develop a deep cascade of CNNs with intermediate data consistency layers which ensure that the originally-sampled k-space in the input is consistent with the reconstruction. Hammernik \etal \cite{hammernik2018learning} develop variational networks to solve reconstruction optimization using gradient descent with CNNs. Similarly, Aggarwal \etal \cite{aggarwal2018modl} use a conjugate gradient algorithm within the reconstruction network. 

In addition to methods operating in the image domain, dual image and k-space methods have also been explored. Eo \etal \cite{eo2018kiki} add an additional k-space reconstruction network to \cite{schlemper2017deep} to enable cross-domain MRI reconstruction. Similarly, Singh \etal \cite{singh2020joint} show that combining frequency and image feature representation learning using two-task-independent layers can improve reconstruction performance over single-domain methods. Zhu \etal \cite{zhu2018image} directly map the undersampled k-space data to its image reconstruction using manifold learning. Moreover, reinforcement learning-aided reconstruction networks were also found to improve the reconstruction quality \cite{wei2020tuning,pineda2020active}. While achieving promising performance, these methods require fully supervised training data from large-scale paired undersampled and fully sampled k-space scans \cite{zbontar2018fastmri,bien2018deep,knoll2020fastmri}. Moreover, these methods only focus on single-contrast MRI reconstruction instead of MC-MRI reconstruction. \\

\noindent\textbf{Self-supervised MRI reconstruction.} Recently, self-supervised reconstruction methods requiring only undersampled k-space data have been proposed for single-contrast MRI reconstruction. HQS-Net~\cite{wang2020neural} decouples the minimization of the data consistency term and regularization term in \cite{schlemper2017deep} based on a neural network, such that network training relies only on undersampled measurements. Yaman \etal \cite{yaman2020self} proposed a physically-guided self-supervised learning method that trains the deep cascade reconstruction network~\cite{schlemper2017deep} by predicting one undersampled k-space data partition using the other data partition, with a similar approach used in Yaman \etal~\cite{yaman2022zeroshot} for subject-specific zero-shot MRI reconstruction. 
\textit{Concurrently} to our work, Korkmaz \etal~\cite{korkmaz2022unsupervised} propose a self-supervised transformer-GAN for zero-shot \textit{instance-specific} optimization and is not comparable to this submission as it focuses on latent noise-to-image GAN mapping and needs to be trained on each new input slice. Hu \etal \cite{hu2021self} also propose to use ISTA-Net~\cite{zhang2018ista} with a parallel training framework for self-supervised single-contrast MRI reconstruction. Furthermore, Zhou \etal \cite{zhou2022dual} devise a triple branch-based dual-domain self-supervised reconstruction framework, achieving promising performance on single-contrast low-field MRI. However, to our knowledge, self-supervised \textit{multi}-constrast MRI reconstruction remains unexplored and is the subject of this work. \\

\noindent\textbf{MC-MRI reconstruction.} Currently, there are few deep learning-based fast MC-MRI reconstruction methods \cite{zhou2020dudornet,xiang2018ultra,sun2019deep,dar2020prior,liu2021deep,liu2021regularization}. Xiang \etal \cite{xiang2018ultra} use fully sampled T1w images as an additional CNN channel input to facilitate accelerated T2w reconstructions. Similarly, Dar \etal \cite{dar2020prior} add adversarial learning and a perceptual loss \cite{johnson2016perceptual} to further improve performance. More recently, Liu \etal \cite{liu2021deep} and Zhou \etal \cite{zhou2020dudornet} feed the fully sampled reference data as an additional channel input into a deep cascade network \cite{schlemper2017deep}. Similar strategies have also been proposed for variational reconstruction~\cite{hammernik2018learning,liu2021regularization}.

\begin{figure}[!htb]
\centering
\includegraphics[width=0.48\textwidth]{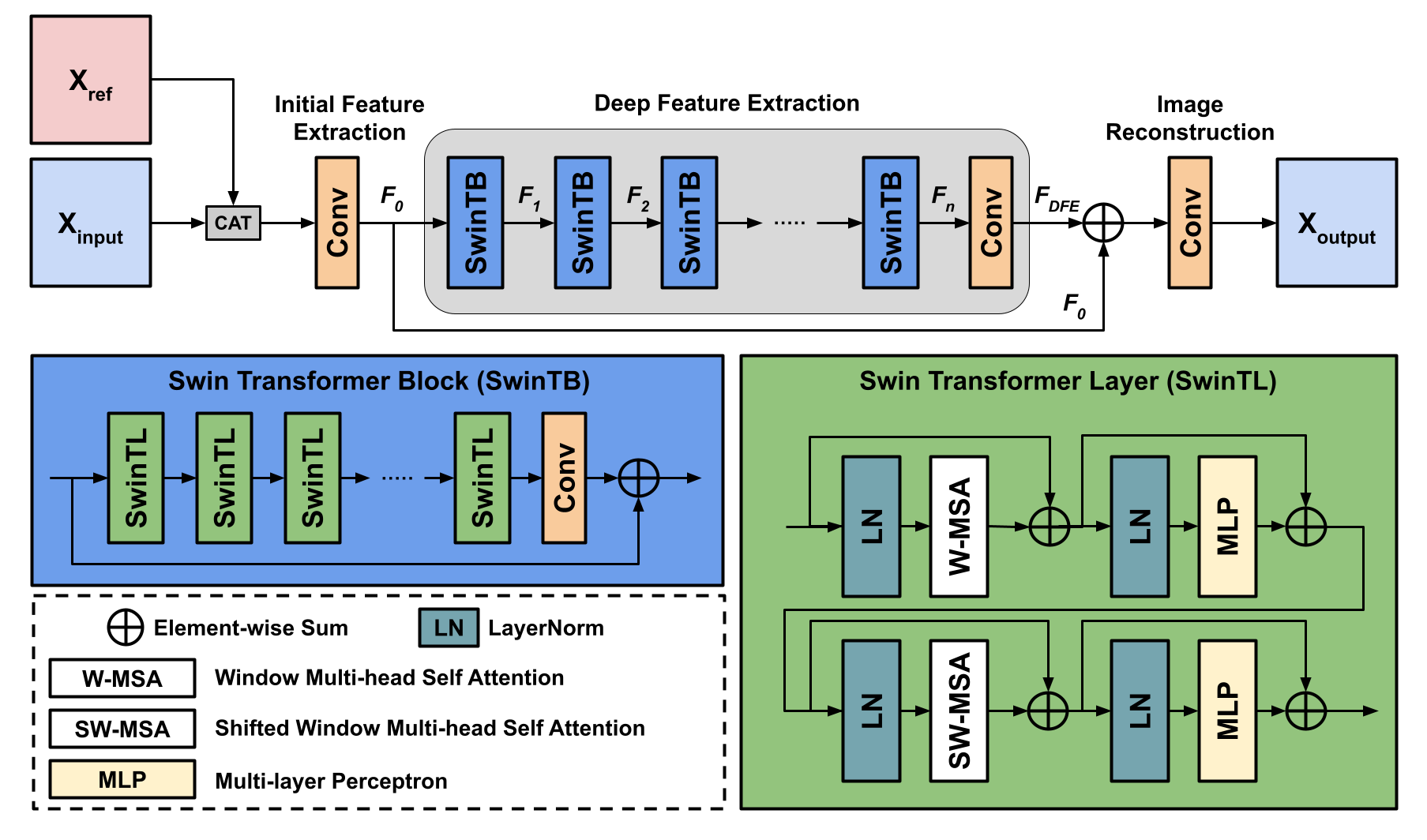}
\caption{The architecture of Swin Transformer Reconstruction Network (SwinRN). It consists of initial feature extraction, deep feature extraction, and image reconstruction modules, and is used as the backbone reconstruction network in Figure \ref{fig:pipline}. }
\label{fig:network}
\end{figure}

%------------------------------------------------------------------------
\section{Methods and Materials}
The overall DSFormer pipeline is illustrated in Figure \ref{fig:pipline} and consists of two major parts: (1) the deep conditional cascade transformer architecture (Fig. \ref{fig:pipline}b) and (2) the dual-domain self-supervised learning strategy used for training DCCT (Fig. \ref{fig:pipline}a).

\subsection{Deep Conditional Cascade Transformer}
DCCT uses a cascaded network design with interleaved data consistency (DC) layers~\cite{schlemper2017deep}. To efficiently exploit multi-contrast information for reconstruction learning, we develop two deep multi-contrast network conditioning mechanisms to better leverage fully-sampled reference acquisitions. To further enable high-quality reconstruction, we propose a Swin Transformer Reconstruction Network (SwinRN) as the backbone network in DCCT. \\

\noindent\textbf{Deep MC-MRI Conditioning.}
With MC-MRI, we use two conditioning methods for sharing reference MRI with target MRI in DCCT: K-space Filling (KF) conditioning and Channel-wise (CC) conditioning. First, we use KF, because multi-contrast MRI depicts distinct physiological properties of imaged tissues, resulting in different image contrast, but the multi-contrast MRI images share the same anatomy. While target contrast zero-padded reconstruction with undersampled k-space data could result in severe artifacts, filling the unacquired k-space with reference contrast k-space data (assuming no motion between the target and reference) can produce alias-reduced reconstruction with the same anatomy and altered contrast. This KF reconstruction can be used as initial DCCT input, so that it can focus on learning contrast conversion instead of de-aliasing. An example of KF is shown in Fig. \ref{fig:kfc}. 

In addition to KF, we also use CC. As illustrated in Figure \ref{fig:pipline}b, the first input to the cascade is the channel-wise concatenation of KF target and the reference contrast MRI image, while the following cascade inputs are the channel-wise concatenations of the previous cascade output and the reference contrast MRI image. \\

\noindent\textbf{Swin Transformer Reconstruction Network.} SwinRN is used as the backbone network for DCCT with its architecture shown in Figure \ref{fig:network}. SwinRN consists of three modules: initial feature extraction (IFE) using a $3 \times 3$ convolutional layer, deep feature extraction (DFE) using multiple Swin Transformer Blocks (SwinTB), and image reconstruction using global residual learning and a $3 \times 3$ convolution layer. The workflow is described as $F_{0} = P_{IFE} ( X_{init} | X_{ref} ),$
where $P_{IFE}$ denotes the IFE operation and $|$ $\cdot$ denotes conditional input. The IFE feature $F_{0}$ is then used for residual learning in the reconstruction step and is fed into multiple SwinTBs for DFE. If there are $n$ SwinTBs, the $n$-th output $F_n$ is $F_{n} = P_{SwinTB_n} ( F_{n-1} )$.

Then, the output of DFE is given by $F_{DFE} = P_{DFE} ( F_{n} )$, where $P_{DFE}$ is a $3 \times 3$ convolutional layer for final feature fusion in DFE. Given $F_{DFE}$ and the global residual connection of $F_{IFE}$, the final reconstruction can be generated via
\begin{equation}
    X_{output} = P_{IR} (F_{DFE} + F_{IFE}),
\end{equation}
where $P_{IR}$ is another $3 \times 3$ convolutional layer for generating a one-channel image reconstruction output. \\

\noindent\textbf{Swin Transformer Block.} 
Each SwinTB (Fig. \ref{fig:network}) consists of multiple Swin transformer layers (SwinTL), a convolution layer for local feature fusion, and a residual connection for local residual learning. Given the input feature $F_{i,0}$ of the $i$-th SwinTB, the intermediate feature is written as:
\begin{equation}
    F_{i,j} = P_{SwinTL_{i,j}} (F_{i, j-1}),
\end{equation}
where $P_{SwinTL_{i,j}}(\cdot)$ is the $j$-th SwinTL in the $i$-th SwinTB. Then, local feature fusion and local residual learning is applied to generate the SwinTB output:
\begin{equation}
    F_{i} = P_{LFF_i}(F_{i,K}) + F_{i,0},
\end{equation}
where $K$ is the number of SwinTL in SwinTB and $P_{LFF_i}$ is a convolutional layer for SwinTB's $i$-th local feature fusion. 

SwinTL \cite{liu2021swin} consists of layer normalization (LN), multi-layer perceptrons (MLP), and multi-head self attention (MSA) modules \cite{vaswani2017attention} with regular windowing (W-MSA) and shifted windowing (SW-MSA) configurations. Given an input with feature size $H \times W \times C$, SwinTL first reshapes the input into $(M \times M) \times (\frac{H}{M} \times \frac{W}{M}) \times C$ by partitioning it into non-overlapping $\frac{H}{M} \times \frac{W}{M}$ windows with each window containing $M \times M$ patches. Then, self-attention can be computed for each window \cite{vaswani2017attention} and can formulate the attention output of W-MSA. To enable cross-window connection, self-attention is also computed for each window by shifting the feature by $(\lfloor\frac{M}{2}\rfloor, \lfloor\frac{M}{2}\rfloor)$ before partitioning. SwinTL processing (Fig. \ref{fig:network}) can thus be summarized as,
\begin{align}
    \bar{z}   & = W \mbox{-} MSA( LN(z) ) + z , \\
    \ddot{z}  & = MLP( LN( \bar{z} ) ) + \bar{z} , \\
    \hat{z}   & = SW \mbox{-} MSA( LN( \ddot{z} ) + \ddot{z}, \\
    \breve{z} & = MLP( LN( \hat{z} ) ) + \hat{z} , 
\end{align}
where $MLP$ is a 2-layer and 30-60 neuron wide MLP with a GELU activation \cite{hendrycks2016gaussian}. 

In summary, SwinRN with SwinTB blocks is embedded in the cascaded framework of DCCT for MRI reconstruction. We use three SwinRNs by default in our cascade, with each SwinRN sharing the same parameters as the default setting. The number of SwinTB in each SwinRN is set to four, with each SwinTB containing four SwinTLs.  

\subsection{Dual-Domain Self-Supervised Learning}
To train DCCT in a self-supervised fashion without using any fully-sampled ground truth data in the target domain, we use dual-domain self-supervision, as illustrated in Figure \ref{fig:pipline}. Let $f_{dcct}(y_{tag}, y_{ref})$ denote DCCT, where $y_{tag}$ is the target contrast's undersampled data and $y_{ref}$ is the reference contrast's fully sampled data. During training, we first randomly partition $y_{tag}$ into two disjoint sets via, \iffalse \\
\noindent\begin{minipage}{.5\linewidth}
\begin{equation}
    y_{p_1} = y_{tag} \odot M_1
\end{equation}
\end{minipage}
\noindent\begin{minipage}{.5\linewidth}
\begin{equation}
    y_{p_2} = y_{tag} \odot M_2
\end{equation}
\end{minipage}\\
\fi
\begin{align}
    y_{p_1} & = y_{tag} \odot M_1 \\
    y_{p_2} & = y_{tag} \odot M_2,
\end{align}
where $\odot$ is element-wise multiplication and $M_1$ and $M_2$ are binary k-space masks for partition 1 and partition 2. Note that $M_1 + M_2 = M_{tag}$, where $M_{tag}$ is the binary mask indicating all under-sampled locations. The partitions $y_{p_1}$ and $y_{p_2}$ are then fed into DCCT for parallel reconstruction,
\begin{align}
    X_{p_1} & = f_{dcct}(y_{p_1}, y_{ref}) \\
    X_{p_2} & = f_{dcct}(y_{p_2}, y_{ref}),
\end{align}
where the networks share the same weights. As the reconstructions of $y_{p_1}$ and $y_{p_2}$ should be consistent with each other, our first loss is an Appearance Consistency (AC) loss operating in the image domain as,
\begin{equation}
    \mathcal{L}_{AC} = \lambda_{1} \mathcal{L}_{img} + \lambda_{2} \mathcal{L}_{grad},
\end{equation}
where,
\begin{equation}
    \mathcal{L}_{img} = ||X_{p_1} - X_{p_2}||_1 
\end{equation}
and,
\begin{align}
    \mathcal{L}_{grad} = ||\nabla_v X_{p_1} - \nabla_v X_{p_2}||_1 + ||\nabla_h X_{p_1} - \nabla_h X_{p_2}||_1,
\end{align}
% \begin{align}
%     \mathcal{L}_{AC} = & \lambda_{1} \mathcal{L}_{img} + \lambda_{2} \mathcal{L}_{grad}, \\
%     \text{where } \mathcal{L}_{img} = & \|X_{p_1} - X_{p_2}\|_1 \ \text{and,} \\
%     \mathcal{L}_{grad} = & \|\nabla_v X_{p_1} - \nabla_v X_{p_2}\|_1 \\ 
%                          & + \|\nabla_h X_{p_1} - \nabla_h X_{p_2}\|_1, \nonumber
% \end{align}
where $\nabla_v$ and $\nabla_h$ are vertical and horiztonal intensity gradient operators, respectively. We empirically found $\lambda_{1} = 1$ and $\lambda_{2} = 0.1$ to achieve optimal performance. 

Our second loss corresponds to a Partition Data Consistency (PDC) loss which operates in k-space. If DCCT can generate a high-quality image from any undersampled k-space measurement, the k-space data of the image predicted from the first partition $y_{p_1}$ should be consistent with the other partition $y_{p_2}$ and vice versa. The predicted k-space partition can be written as,
\begin{align}
    y_{2 \rightarrow 1} & = \mathcal{F} (X_{p_2}) \odot M_1 \\
    y_{1 \rightarrow 2} & = \mathcal{F} (X_{p_1}) \odot M_2,
\end{align}
Therefore, the PDC loss is formulated as,
\begin{equation}
    \mathcal{L}_{PDC} = ||y_{2 \rightarrow 1} - y_{p_1}||_1 + ||y_{1 \rightarrow 2} - y_{p_2}||_1,
\end{equation}
where the first and second term are the partial data consistency losses for partitions 1 and 2, respectively. Combining the AC loss in the image domain and the PDC loss in k-space, our total loss can be written as,
\begin{equation}
    \mathcal{L}_{tot} = \mathcal{L}_{AC} + \lambda_{3} \mathcal{L}_{PDC}
\end{equation}
where $\lambda_{3}=0.1$ is used to balance the scale between k-space and image domain losses. 

\begin{table*} [tb!]
\footnotesize
\centering
\caption{Quantitative comparison of T2 (left sub-table) and PD (right sub-table) reconstructions under three different acceleration settings for the target contrast MRI. Fully supervised methods and self-supervised methods are marked in \textbf{bold} and \underline{underlined}, respectively. Best results are marked in \textcolor{red}{red}.}
\label{tab:comp_methods}
    %\begin{tabular}{l|c|c|c||c|c|c}
    \begin{tabularx}{0.999\textwidth}{l@{\hskip 0.1in}c@{\hskip 0.1in}c@{\hskip 0.1in}c@{\hskip 0.1in}c@{\hskip 0.1in}c@{\hskip 0.1in}c@{\hskip 0.1in}c@{\hskip 0.1in}c@{\hskip 0.1in}c@{\hskip 0.1in}c}
        \toprule
        \textbf{PSNR/SSIM}                              & \multicolumn{4}{c}{\textbf{Target: T2w} $\mid$ \textbf{Reference: PD}}       &  \multicolumn{4}{c}{\textbf{Target:PD} $\mid$ \textbf{Reference:T2w}}       & \textbf{Runtime}     &  \textbf{Number}    \\%\Tstrut\Bstrut\\
        \cline{2-9}                                                                                                                                                                                       
        \textbf{Methods}                                & $\times 2$      & $\times 4$       & $\times 6$     & $\times 8$             & $\times 2$      & $\times 4$       & $\times 6$     & $\times 8$            & \textbf{(ms)}        & \textbf{of Param}  \\%\Tstrut\Bstrut\\
        \midrule
        Zero-padding                                    & $24.86/.761$    & $22.72/.679$     & $21.48/.623$   & $19.11/.597$           & $23.92/.744$    & $21.68/.663$     & $20.47/.610$   & $18.78/.579$          & - & - \\%\Tstrut\Bstrut\\
        % \hline
        CS-TV\cite{huang2014fast}                       & $30.18/.890$    & $29.04/.862$     & $26.13/.789$   & $25.67/.762$           & $30.09/.887$    & $29.18/.861$     & $26.05/.787$   & $25.21/.750$          & $3023.8$ & - \\%\Tstrut\Bstrut\\
        \midrule
        \textbf{SSDU}\cite{yaman2020self}               & $42.54/.981$    & $38.47/.976$     & $34.21/.972$   & $31.72/.961$           & $41.78/.983$    & $37.63/.979$     & $33.99/.968$   & $31.61/.960$          & $49.9$ & $0.11M$ \\%\Tstrut\Bstrut\\
        % \hline
        \textbf{HQSNet}\cite{wang2020neural}            & $40.13/.980$    & $37.79/.973$     & $33.38/.969$   & $31.12/.958$           & $40.82/.981$    & $35.97/.974$     & $33.37/.967$   & $31.03/.956$          & $49.9$ & $0.11M$  \\%\Tstrut\Bstrut\\
        % \hline
        \textbf{SSISTA}\cite{hu2021self}                & $42.09/.980$    & $38.29/.974$     & $34.13/.971$   & $31.51/.960$           & $41.29/.982$    & $37.43/.977$     & $33.67/.968$   & $31.43/.959$          & $40.8$ & $0.38M$ \\%\Tstrut\Bstrut\\
        \midrule
        \underline{UFNet}\cite{xiang2018ultra}          & $32.30/.970$    & $32.07/.969$     & $31.85/.967$   & $30.02/.950$           & $32.22/.971$    & $32.06/.969$     & $31.88/.968$   & $29.99/.948$          & $10.6$ & $7.6M$ \\%\Tstrut\Bstrut\\
        %\hline
        \underline{VarNet}\cite{liu2021regularization}  & $33.01/.973$    & $32.71/.971$     & $32.43/.970$   & $30.65/.957$           & $33.08/.974$    & $32.86/.972$     & $32.67/.970$   & $30.55/.955$          & $48.7$ & $8.2M$  \\%\Tstrut\Bstrut\\
        %\hline
        \underline{MCNet}\cite{liu2021deep}             & $43.79/.989$    & $39.14/.983$     & $35.61/.972$   & $32.12/.963$           & $42.90/.988$    & $38.56/.979$     & $35.45/.971$   & $32.03/.961$          & $49.9$ & $0.11M$  \\%\Tstrut\Bstrut\\
        \midrule
        \textbf{DSFormer}                               & \textcolor{red}{$45.05/.993$}    & \textcolor{red}{$40.31/.985$}     & \textcolor{red}{$37.04/.977$}   & \textcolor{red}{$33.65/.969$}  & \textcolor{red}{$45.07/.993$}    & \textcolor{red}{$40.52/.987$}     & \textcolor{red}{$37.45/.982$}    & \textcolor{red}{$33.48/.967$}        & $51.3$ & $0.18M$            \\%\Tstrut\Bstrut\\
        \bottomrule
    \end{tabularx}
\end{table*}

\begin{figure*}[!htb]
\centering
\includegraphics[width=0.94\textwidth]{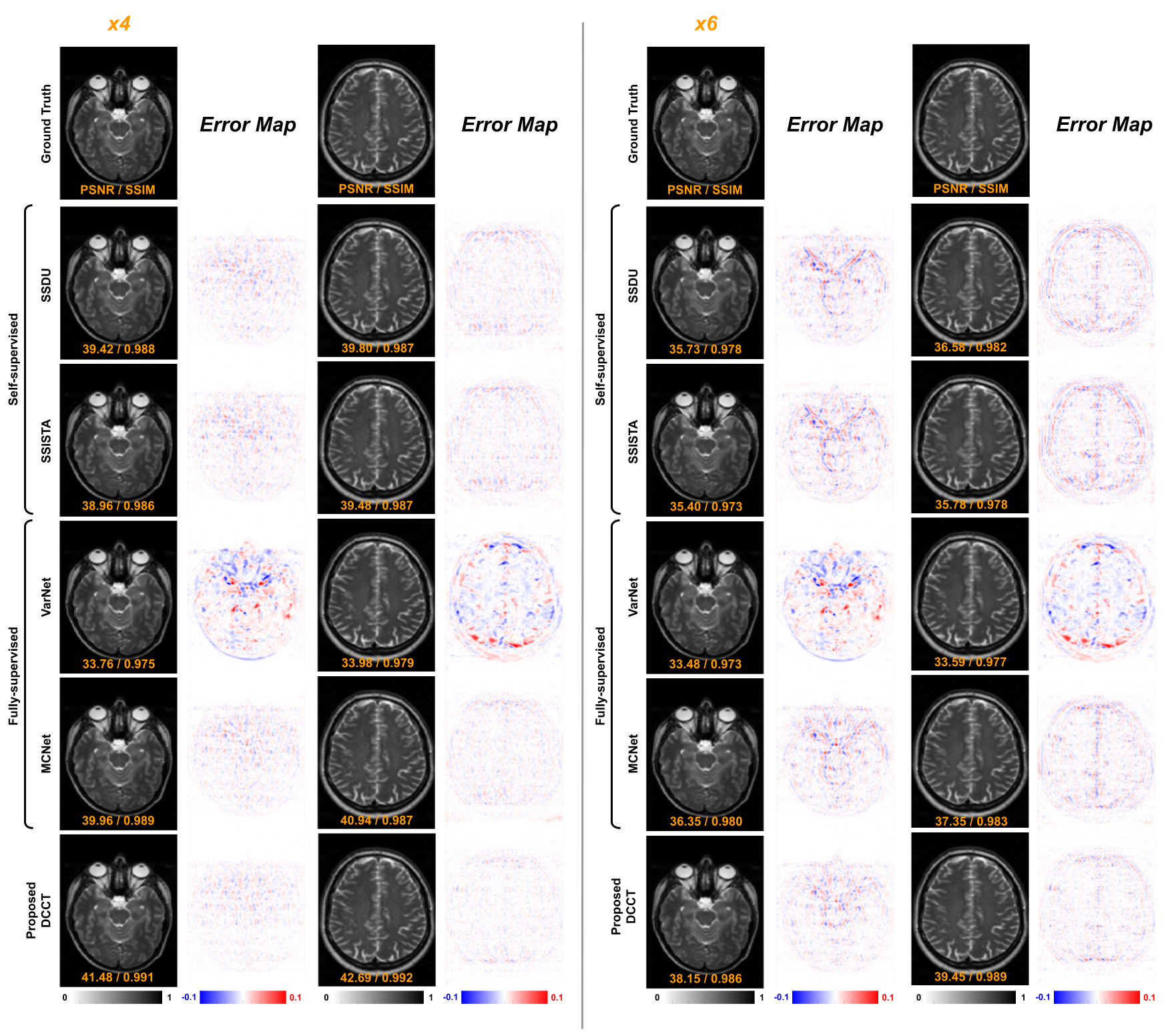}
\caption{Qualitative comparisons of T2 reconstructions using $\times 4$ and $\times 6$ acceleration. For T2 reconstruction, PD is used as the reference contrast. The corresponding error maps between ground truth images and the reconstructions are illustrated in BWR colormaps. Across both supervised and self-supervised methods, DSFormer achieves the highest-fidelity reconstructions due to its improved architecture, dual-domain self-supervision, and conditioning mechanisms.}
\label{fig:comp_methods_T2}
\end{figure*}

\subsection{Data Preparation} 
We use 578 MC-MRI subjects with both T2-weighted and Proton Density (PD)-weighted acquisitions from IXI\footnote{https://brain-development.org/ixi-dataset/, CC BY-SA 3.0 license} for our experiments. The registered MC-MRI data consisting of 11808 pairs of T2 and PD weighted axial slices were split subject-wise into 8376 pairs for training, 1080 for validation, and 2352 for testing, with no slices from any subject overlapping. We consider two MC-MRI scenarios in our experiments: accelerating T2-weighted acquisition (the target protocol) by utilizing a fully sampled PD-weighted acquisition (the reference protocol), and accelerating PD-weighted target acquisition with a fully sampled T2-weighted reference. Here, we consider the Cartesian sampling pattern with the acceleration factor (R) set to a value between 2 and 8 corresponding to acceleration in acquisition time for the target protocol.

\subsection{Evaluation Metrics and Baselines Comparisons} 
Benchmark results are presented on 2352 test slices from 114 patients. We evaluate the target reconstruction results using Peak Signal-to-Noise Ratio (PSNR) and Structural Similarity Index (SSIM) computed against ground truth. For baseline comparison, we first compare our results against previous fully-supervised MC-MRI reconstruction methods that require ground truth fully-sampled data, including UFNet~\cite{xiang2018ultra}, MCNet~\cite{liu2021deep}, and VarNet~\cite{liu2021regularization}. To further benchmark against previous self-supervised MRI reconstruction methods originally designed for single-contrasts, we also extend SSDU~\cite{yaman2020self}, HQSNet~\cite{wang2020neural}, and SSISTA~\cite{hu2021self} to the MC-MRI setting by using the reference contrast as an extra input channel. As an upper bound, we also compare self-supervised DSFormer against a supervised-variant where the DCCT of DSFormer was trained in a fully supervised fashion with ground truth available.

\subsection{Implementation Details}
We implement our method in Pytorch and perform experiments using an NVIDIA Quadro RTX 8000 GPU with 48GB memory. The Adam solver \cite{kingma2014adam} was used to optimize our models with $lr = 2 \times 10^{-4}$, $\beta_{1} = 0.9$, and $\beta_{2} = 0.999$. We use a batch size of 3 during training. In DSFormer, the number of cascades can be flexibly adjusted and is set to three as the default setting in the main experiments and is swept over in Fig. \ref{fig:plot_n_cascade}. The SwinRN shares the same parameters in each cascade. The number of SwinTB in each SwinRN is set to four, with each SwinTB containing four SwinTLs. During training, the data partitioning rate is randomly generated between $[0.2, 0.8]$ on-the-fly which separates the undersampled k-space data into two disjoint k-space data and augments the training data. For baseline implementations, we use the official code releases of SSDU\footnote{https://github.com/byaman14/SSDU}, HQSNet\footnote{https://github.com/alanqrwang/HQSNet}, and SSISTA\footnote{https://github.com/chenhu96/Self-Supervised-MRI-Reconstruction}, and implement UFNet, MCNet, and VarNet in-house due to unavailability of official code. The hyperparameters of each method are tuned on the validation set with test data held-out for final evaluation.

%------------------------------------------------------------------------
\section{Experimental Results}
\subsection{Image Quality Evaluation and Comparison}
Quantitative evaluations on two different MC-MRI scenarios under three different acceleration settings are summarized in Table \ref{tab:comp_methods}. The left sub-table summarizes MC-MRI reconstruction with T2 target contrast and PD reference contrast (T2 reconstructions were evaluated here). Among fully supervised methods, MCNet~\cite{liu2021deep} achieves the best T2 reconstruction performance with PSNR up to $43.79$ dB and SSIM up to $0.989$ when using $\times 2$ acceleration. It can also be observed that MCNet~\cite{liu2021deep} consistently outperforms the previous self-supervised MRI reconstruction methods modified to operate on multi-contrast data. In the last row of Table \ref{tab:comp_methods}, we see that DSFormer trained with self-supervision alone outperforms supervised baselines and increases PSNR from $43.79$ dB to $45.05$ dB and SSIM from $0.989$ to $0.993$. Similar observations are made for the $\times 4$ accelerated T2 experiments where DSFormer outperforms MCNet, with PSNR increasing from $39.14$ dB to $40.31$ dB and SSIM increasing from $0.983$ to $0.985$. 

As expected, the reconstruction performance of all methods decreases as the acceleration rate increases. However, DSFormer is still able to widely outperform previous supervised methods and keep PSNR at $37.04$ and SSIM at $0.977$ with $\times 6$ accelerated T2 reconstruction. The inference run time and the number of model parameters of different methods are also summarized in Table \ref{tab:comp_methods}, with the deep learning methods achieving orders of magnitude faster reconstruction over iterative methods like CS-TV~\cite{huang2014fast}. As compared to the previous best results of MCNet, DSFormer requires only a slightly increased number of parameters and run-time to achieve improved reconstruction performance.

The qualitative comparison of various T2 reconstructions is shown in Fig. \ref{fig:comp_methods_T2}, illustrating $\times 4$ and $\times 6$ acceleration settings. Reconstructions with zero padding create significant aliasing artifacts and lose anatomical details. While both VarNet \cite{liu2021regularization} and MCNet \cite{liu2021deep} significantly reduce the aliasing artifacts with decreased reconstruction error, they require fully supervised training from paired data. On the other hand, DSFormer, using only self-supervision and multi-contrast conditioning, further reduces the residual error between the reconstruction and the ground truth and yields superior reconstruction quality. The full qualitative comparison of T2 reconstruction is visualized in Figure S1 in the supplemental materials.

The right sub-table of Table \ref{tab:comp_methods} summarizes MC-MRI with PD target contrast and T2 reference contrast, where PD reconstructions were evaluated. Similar observations are made for PD reconstruction, where DSFormer still achieves the best reconstruction under all three acceleration settings over previous fully supervised methods~\cite{xiang2018ultra,liu2021regularization,liu2021deep} and modified previous self-supervised methods~\cite{yaman2020self,wang2020neural,hu2021self}. The qualitative comparison of several PD reconstruction methods is shown in the supplementary materials.

\begin{figure}[!htb]
\centering
\includegraphics[width=0.48\textwidth]{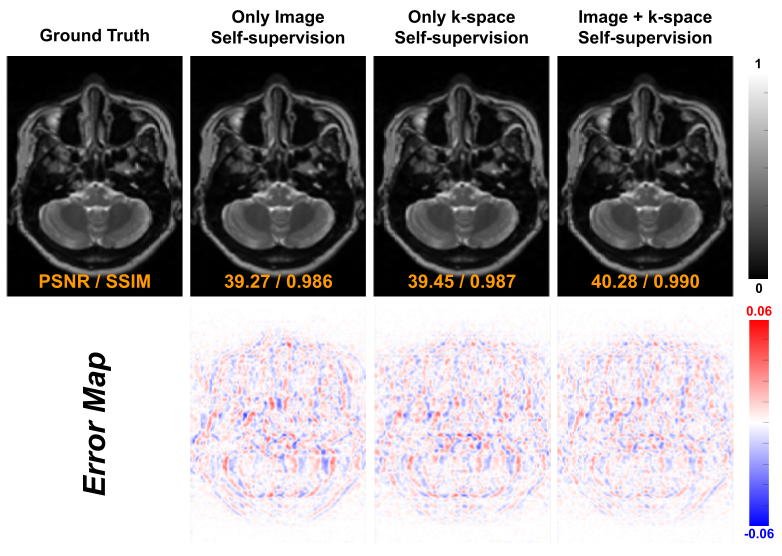}
\caption{Further ablation studies comparing MRI reconstruction from different self-supervision settings. Lower error is better. }
\label{fig:comp_domains_T2}
\end{figure}

\subsection{Ablation Studies}
\noindent\textbf{Dual-domain self-supervision}. To isolate the individual utility of the various components of dual-domain self-supervision, we evaluate performance using either only image-domain self-supervision ($\mathcal{L}_{AC}$) or only k-space self-supervision ($\mathcal{L}_{PDC}$). The quantitative comparison is summarized in Table \ref{tab:comp_domains}. We observe that using either only k-space self-supervision or only image-domain self-supervision still yields strong reconstruction quality, with a PSNR of $39.95$ under $\times 4$ acceleration, which indicates self-supervision in both domains can help with reconstruction. Combining both image-domain and k-space self-supervision yields the best reconstruction performance. A visual comparison of reconstructions from different self-supervision settings is illustrated in Figure \ref{fig:comp_domains_T2}. 

\begin{table} [htb!]
\footnotesize
\centering
\caption{Quantitative comparison of T2 reconstruction performance from different self-supervision settings, including image-domain only self-supervision, k-space only self-supervision, and the proposed dual-domain self-supervision. Higher is better.}
\label{tab:comp_domains}
    \begin{tabular}{l@{\hskip 0.25in}c@{\hskip 0.1in}c@{\hskip 0.1in}c}
        \hline
        \textbf{PSNR/SSIM}              & $\times 4$        & $\times 6$        \Tstrut\Bstrut\\
        \hline
        Only k-space self-supervision   & $40.23/0.982$     & $36.89/0.975$     \Tstrut\Bstrut\\
        % \hline
        Only image self-supervision     & $39.95/0.981$     & $36.55/0.973$     \Tstrut\Bstrut\\
        \hline
        DSFormer (proposed)             & $40.31/0.985$     & $37.04/0.977$     \Tstrut\Bstrut\\
        \hline
    \end{tabular}
\end{table}

\noindent\textbf{Deep MC-MRI conditioning}. To understand the impact of KF and CC at the initial network input, we evaluate DSFormer performance with or without KF and CC, with results summarized in Table \ref{tab:comp_abl}. DSFormer without both KF and CC implies no multi-contrast data is used and results in only $38.24$ PSNR under $R=4$ setting. Under the same setting, DSFormer with either KF-only or CC-only can integrate multi-contrast information and improves PSNR from $38.24$ to $39.06$ with KF-only and $40.22$ with CC-only. Using both KF and CC leads to the best performance where PSNR is further improved to $40.31$. Similar trends are observed for $6\times$ acceleration.

\begin{table} [htb!]
\footnotesize
\centering
\caption{Quantitative comparison of T2 reconstruction performance when using DSFormer with or without KF and CC and when training DSFormer in a fully supervised manner similar to \cite{liu2021deep,xiang2018ultra} without our proposed consistency losses. $\dagger$ means fully supervised training.}
\label{tab:comp_abl}
    \begin{tabular}{l@{\hskip 0.25in}c@{\hskip 0.25in}c@{\hskip 0.25in}c}
        \hline
        \textbf{PSNR/SSIM}              & $\times 4$        & $\times 6$        \Tstrut\Bstrut\\
        \hline
        DSFormer w/o KF                 & $40.22/0.982$     & $36.93/0.972$     \Tstrut\Bstrut\\
        % \hline
        DSFormer w/o CC                 & $39.06/0.979$     & $35.22/0.970$     \Tstrut\Bstrut\\
        % \hline
        DSFormer w/o KF and CC          & $38.24/0.971$     & $34.51/0.961$     \Tstrut\Bstrut\\
        % \hline
        DSFormer                        & $40.31/0.985$     & $37.04/0.977$     \Tstrut\Bstrut\\
        \hline
        \hline
        $\dagger$DSFormer (Upper Bound)    & $40.34/0.989$     & $37.12/0.981$     \Tstrut\Bstrut\\
        % \hline
        $\dagger$DSFormer w/o KF           & $40.25/0.983$     & $36.99/0.974$     \Tstrut\Bstrut\\
        % \hline
        $\dagger$DSFormer w/o CC           & $39.12/0.981$     & $35.31/0.972$     \Tstrut\Bstrut\\
        % \hline
        $\dagger$DSFormer w/o KF and CC    & $38.42/0.976$     & $34.70/0.965$     \Tstrut\Bstrut\\
        \hline
    \end{tabular}
\end{table}

\noindent\textbf{Fully-supervised vs. Self-supervised DSFormer}. 
%As DSFormer is self-supervised with no paired data used in training, 
In order to understand the performance gap between full supervision and dual-domain self-supervision, we compare the reconstruction performance of DCCT trained without partitioning the input and replacing its consistency losses with direct image-domain reconstruction losses against the target ground truth, similar to the supervised training in \cite{xiang2018ultra,liu2021deep}. Quantitative comparisons are summarized in Table S1. As an upper bound, fully supervised DSFormer achieved reconstruction PSNR of $40.34$ dB under $\times 4$ acceleration, which is only $\sim 0.03$ higher than self-supervised DSFormer in terms of PSNR. SwinRN effectiveness can be further evaluated by comparing supervised DSFormer w/o KF with MCNet (Table \ref{tab:comp_methods}) where both methods share the same cascade framework except the difference in the backbone network. We can observe our supervised method based on SwinRN achieves PSNR of $40.25$ which is significantly better than MCNet based on simple sequential convolutional layers with residual connection with PSNR of $39.14$ under $\times4$ setting.
However, at $6\times$ acceleration we see a gap emerging between supervised DSFormer and self-supervised DSFormer, each achieving 37.12 and 37.04 dB PSNR, respectively, indicating that there is still a performance benefit to using fully sampled training data under higher acceleration factors.

\begin{figure}[!htb]
\centering
\includegraphics[width=0.48\textwidth]{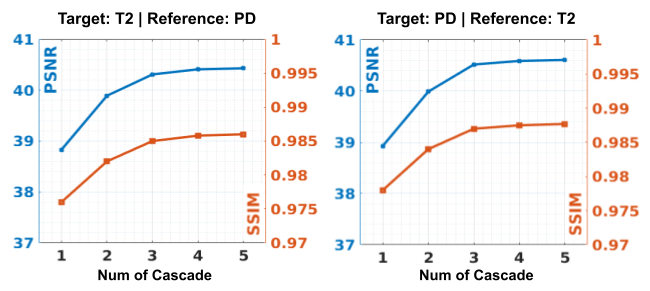}
\caption{The effect of increasing the number of cascaded SwinRNs in DSFormer at $\times 4$ acceleration.}
\label{fig:plot_n_cascade}
\end{figure}

\noindent\textbf{Impact of the number of cascades}. As the number of cascades can be flexibly adjusted in DSFormer, we analyze the effect of increasing the number of cascaded blocks in our framework, with the result summarized in Figure \ref{fig:plot_n_cascade} using $\times4$ acceleration. Using a higher number of cascaded blocks boosts the reconstruction performance, with gains asymptotically stabilizing on further increases beyond three blocks. In T2 reconstruction, increasing the number of cascaded blocks from 3 to 4 only increases PSNR by less than 0.002 dB. Similar observations can be made from the PD reconstructions.

%------------------------------------------------------------------------
\section{Conclusion}
We developed DSFormer, a dual-domain self-supervised transformer for accelerated multi-contrast MRI reconstruction. DSFormer proposed a deep conditional cascaded transformer architecture trained under both k-space and image domain self-supervision. Benchmarks against established baselines demonstrate that DSFormer outperformed previous \textit{fully supervised methods} that require training with paired data (Table \ref{tab:comp_methods}) and that DSFormer achieves nearly the same performance when trained with either full supervision or with our proposed dual-domain self-supervision (Table \ref{tab:comp_abl}), almost closing the gap between supervised and self-supervised methods for accelerated MRI reconstruction.

% \clearpage
{\small
\bibliographystyle{ieee_fullname}
\bibliography{egbib}
}

\end{document}